\documentclass[aps,prb,reprint,groupedaddress]{revtex4-1}
\usepackage{graphicx}  % needed for figures (not in this templat)

\begin{document}

%\preprint{}

\title{
Quantitative Analysis of the Disorder Broadening and the Intrinsic Gap \\ 
  for the $\nu=5/2$ Fractional Quantum Hall State}
%Disorder Broadening and the Fractional Quantum Hall States of the Second Landau Level}
% Supression of the Intrinsic gap by disorder, inhibition

\author{N. Samkharadze$^1$, J.D. Watson$^{1,2}$, G. Gardner$^2$, M.J. Manfra$^{1,2,3}$ , L.N. Pfeiffer$^4$, K.W. West$^4$, 
        and G.A. Cs\'{a}thy$^1$ \footnote{gcsathy@purdue.edu}}

\affiliation{${}^1$ Department of Physics, Purdue University, West Lafayette, IN 47907, USA \\
${}^2$ Birck Nanotechnology Center Purdue University, West Lafayette, IN 47907, USA \\
${}^3$ School of Materials Engineering and School of Electrical and Computer Engineering,
Purdue University, West Lafayette, IN 47907, USA \\
${}^4$Department of Electrical Engineering, Princeton University, Princeton, NJ 08544\\} 


\date{\today}

\begin{abstract}

We report a reliable method to estimate the disorder broadening parameter 
from the scaling of the gaps of the even and major odd denominator fractional quantum Hall states
of the second Landau level. We apply this technique to several samples 
of vastly different densities and grown in different MBE chambers. 
Excellent agreement is found between the estimated intrinsic and numerically obtained
energy gaps for the $\nu=5/2$ fractional quantum Hall state.
Futhermore, we quantify, for the first time, the dependence of the intrinsic gap at $\nu=5/2$ 
on Landau level mixing.

\end{abstract}

\maketitle

Disorder plays an important role in the formation of the
fractional quantum Hall states (FQHS) observed in the two-dimensional electron gas (2DEG) \cite{tsui,kun03}. 
% subjected to a perpendicular magnetic field $B$ 
While qualitative aspects of the effect of the disorder have been appreciated early on \cite{tsui}, the quantitative 
effect of the disorder on the properties such as the energy gap of the FQHSs remains poorly understood.

Currently significant effort is focused on the FQHS at the Landau level (LL) filling factor $\nu=5/2$
\cite{willett87,pan99,eisen02,csa10,xia04,miller07,radu08,dolev08,willett10,bid10,vivek11,dean08,choi08,xia10,nuebler10,pan08,umansky09,gamez11,pan11,dolev11}. 
This state does not belong to the sequence of FQHS described by the theory of weakly interacting
composite fermions (CF) \cite{jainCF,halp93} and, therefore it may have exotic quantum correlations which 
are not of the Laughlin type \cite{laughlin,jainCF}. 
It is believed that the $\nu=5/2$ FQHS arises from a $p$-wave pairing of the CF
described by either the Pfaffian \cite{moore91,kun06,moller08,storni10} 
or the anti-Pfaffian \cite{levin07,lee07,rezayi11} wavefunction. 
%The Pfaffian state is of considerable interest since its quasiparticles are predicted to obey exotic
%non-Abelian braiding statistics and that may be used for error-free topological
%quantum computing \cite{dassarma05}. 

Agreement between the measured energy gap $\Delta^{meas}_{5/2}$ and that from numerical studies 
is a necessary condition for an identification of the $\nu=5/2$ FQHS with the Pfaffian 
\cite{feinguin08,morf98,morf03,wojs06,wojs10,park98,morf02,peters08,papic09}.
Gaps in numerical studies are always calculated in the absence of any disorder and they
must be therefore compared to the measured gaps extrapolated to zero disorder, 
also called the intrinsic gap $\Delta^{int}$.
%The gap suppression by the disorder is small for the most prominent FQHS at $\nu=1/3$ \cite{willett88}.
%In contrast, $\Delta^{meas}_{5/2}$
%can be a small fraction of its intrinsic gap $\Delta^{int}_{5/2}$ for the $\nu=5/2$ FQHS \cite{pan99}. 
While the effect of disorder on the gap is small for the most prominent FQHS at $\nu=1/3$ \cite{willett88},
it is quite large at $\nu=5/2$ \cite{pan99}. 
Hence a quantitative knowledge of the gap suppression by the disorder and of the intrinsic gap play 
a significant role in the identification of the nature of the exotic FQHSs in the second LL.

Three different methods have been used so far to obtain the intrinsic gap at $\nu=5/2$  but, to date, they have not yielded consistent results.
A scaling of the measured gaps of the even denominator FQHS \cite{morf03} and
an estimation using the quantum lifetime \cite{choi08}
found good agreement between experimental and numerical intrinsic gaps.
However, extrapolations of $\Delta^{meas}_{5/2}$ to infinite mobility \cite{pan08,dean08}
%%  error1
and a recent estimation from the quantum lifetime \cite{nuebler10}
found an intrinsic gap three times smaller than expected.
%Futhermore, an analysis of the gaps in tilted magnetic fields calls the Pfaffian picture into
%question by favoring a spin-unpolarized state at $\nu=5/2$ 
%\cite{dassarma10}. 
This situation calls for a reexamination of the extraction of $\Delta^{int}_{5/2}$ 
from the measurements.
% and of the comparison of the experimentally and theoretically obtained gaps.

We adopt the method of quantifying the effect of the disorder
using the even denominator FQHS 
\cite{morf03} and propose a new method using the two odd denominator FQHS at $\nu=2+1/3$ and $2+2/3$.
We find that, within experimental error, these two methods give consistent results
in samples of very different densities and grown in different MBE chambers.
The intrinsic gaps $\Delta^{int}_{5/2}$ found are
in excellent agreement with gaps calculated from numerics which include the effects of Landau level mixing (LLM)
and finite extent of the wavefunction.  Our results strongly indicate that the paired-state Pfaffian is the correct description of the $\nu=5/2$ FQHS.
We also show that $\Delta^{int}_{5/2}$ cannot be reliably obtained from the quantum lifetime or from extrapolation
of $\Delta^{meas}_{5/2}$ to infinite mobility.
From the dependence of the intrinsic gap of the $\nu=5/2$ FQHS on LLM 
we find that the $\nu=5/2$ FQHS becomes unstable beyond a threshold value of the LLM parameter $\kappa_{th}=2.9$.  
% and extrapolate it to the limit of zero LLM. 
% We conclude that theory seems to overestimate the gap reduction effects of LLM and we establish an 
% unexpected scaling of $\Gamma$ with the density.

There are two GaAs quantum well samples used in this study.  
Sample $A$ grown at Princeton has a well width of $56$nm, a density 
$n=8.30 \times 10^{10} $cm$^{-2}$, and mobility $\mu = 12 \times 10^{6}$cm$^{2}$/Vs.
Sample $B$ grown in a newly built MBE chamber at Purdue has a width of 30nm,
a density $n=2.78 \times 10^{11} $cm$^{-2}$ and mobility $\mu = 11 \times 10^{6}$cm$^{2}$/Vs.
Both wells are flanked by Al$_{0.24}$Ga$_{0.76}$As barriers with
the Si donors placed symmetrically from the well at 320nm and 78nm, respectively. 
Samples are mounted in a He$^3$ immersion cell described in detail in Ref.\cite{csa11}.

\begin{figure}[t]
% \begin{center}
 \includegraphics[width=1\columnwidth]{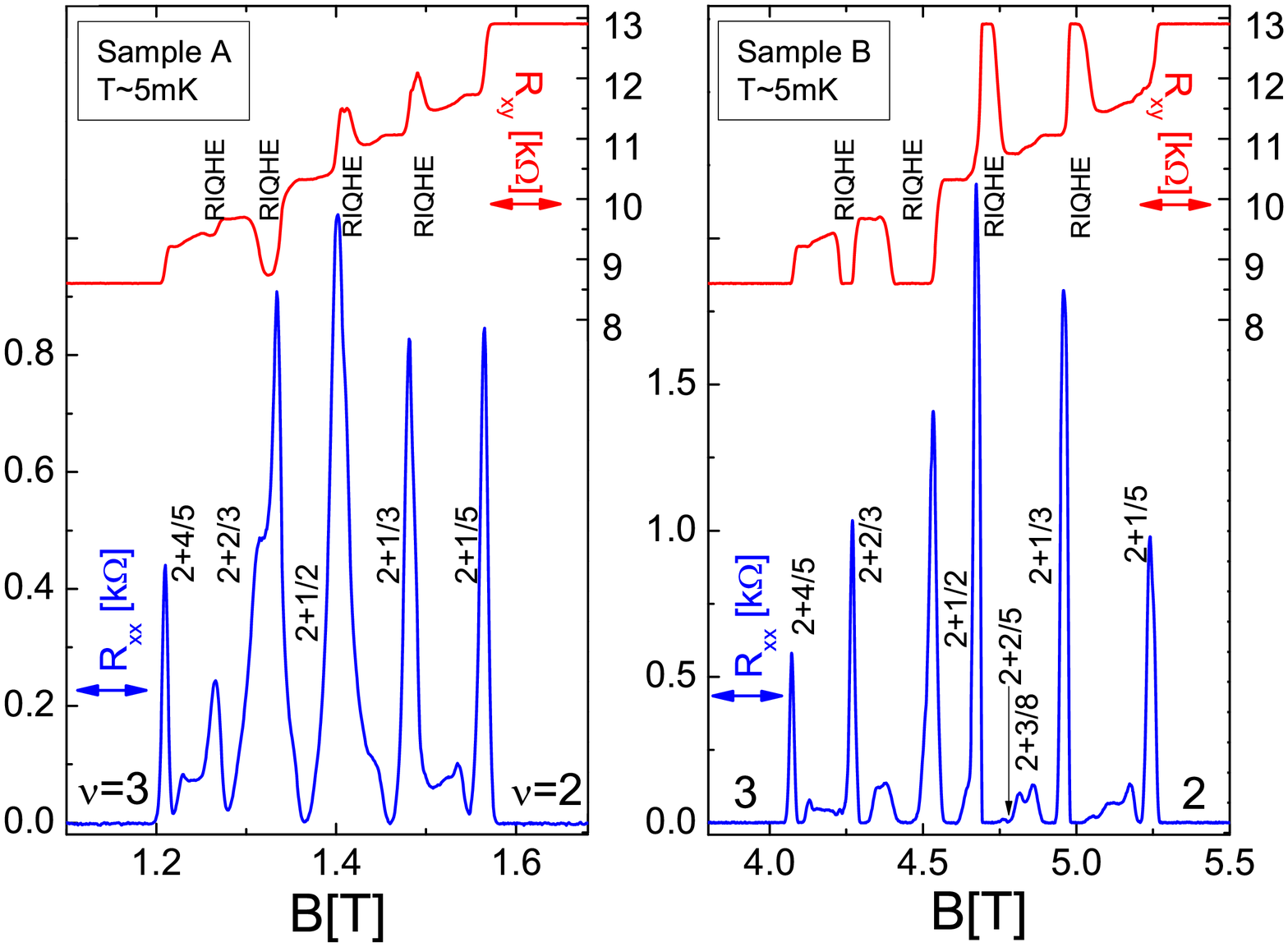}
% \end{center}
 \caption{\label{f1}
Magnetotransport data in the region of the upper spin branch of the second Landau level.
We mark the filling factors $\nu$ of the observed FQHSs and the reentrant integer quantum Hall states (RIQHS).
Note the vastly different densities of the two samples.
}
\end{figure}

Fig. 1 shows the longitudinal $R_{xx}$ and transverse $R_{xy}$ resistances as function of the magnetic field $B$
in the second LL (i.e. for $2<\nu<3$)
for the two samples. The $\nu=5/2$ FQHS is fully quantized in both samples; this state in sample $A$
occurs at the lowest magnetic field of 1.37T yet reported \cite{dean08,xia10,nuebler10}. 
Other FQHS also develop. Notably, sample $B$ has a fully quantized $2+2/5$ FQHS 
and an incipient $2+3/8$ FQHS, hallmarks of the highest quality samples \cite{xia04,csa10}.
We note that the mobility of sample $B$ is about a factor of 3 lower
than that of other samples exhibiting similar higher order FQHSs \cite{xia04,csa10}. 

Fig. 2 shows the Arrhenius plots of $R_{xx}$ for selected FQHS observed in the second LL of sample $A$. 
The $\Delta^{meas}$ extracted using $R_{xx} \! \propto \! \exp \left( -\Delta^{meas}/2T \right) $
are shown in Table I. 
%Similarly to samples of higher densities, the gap at $\nu=2+2/3$ 
%is about a factor of two lower than that at $\nu=2+1/3$.  
Since in this work we will analyze the gaps of the $\nu=5/2$, $7/2$, $2+1/3$, and $2+2/3$ FQHSs,
in Table I. we also consider samples for which the gaps for these four FQHSs are available
\cite{dean08,csa10}. For the sample in Ref.\cite{csa10} $\Delta^{meas}_{7/2}=240$mK.

In order to estimate the intrinsic gap $\Delta^{int}_{5/2}$ for the $\nu=5/2$ FQHS, an extrapolation of $\Delta^{meas}_{5/2}$
to infinite mobility has recently been used \cite{dean08,pan08}. We argue that such an extrapolation 
is inherently inaccurate. Indeed, our sample $B$ shows unusually large gaps in spite of a modest mobility $\mu = 11 \times 10^{6}$cm$^{2}$/Vs
and, therefore, it is quite a bit off from the extrapolation done in Refs.\cite{dean08,pan08}.
We conclude that, as previously noted \cite{nuebler10,pan08,umansky09,gamez11,pan11}, the intrinsic gap
does not directly correlate with the mobility. 
%
%First, there is a large spread of $\Delta^{meas}$ in different 
%samples of similar $\mu$. Second, the functional dependence of $\Delta^{meas}_{5/2}(1/\mu)$ is not known. Third,
%$\Delta^{meas}_{5/2}$ is reported in several recent experimental work 
%to be affected differently by different types of disorder
%and, therefore, it is not likely to have a universal and/or monotonic dependence on
%the mobility \cite{umansky09,gamez11}. 
%We conclude that the mobility is not necessarily a good measure of the effect of the disorder on the energy gaps of FQHS.
%This conclusion is further strengthened by our data on sample $B$ for which we measured strikingly high 
%gaps in spite of a modest mobility $\mu = 11 \times 10^{6}$cm$^{2}$/Vs.

\begin{figure}[t]
% \begin{center}
 \includegraphics[width=0.9\columnwidth]{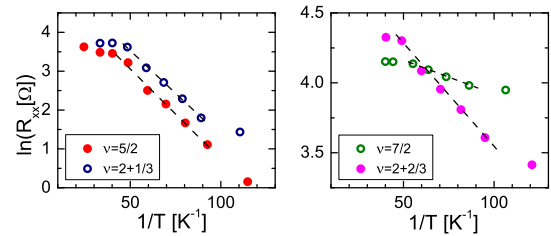}
% \end{center}
 \caption{\label{f2}
 Arrhenius plots used for the extraction of the activation gaps in sample $A$. 
}
\end{figure}

\begin{table}[b]
\caption{Energy gaps $\Delta^{meas}$ in units of mK for our samples.}
\begin{ruledtabular}
\begin{tabular}{l c c c c}
sample & $\Delta^{meas}_{5/2}$ & $\Delta^{meas}_{7/2}$ & $\Delta^{meas}_{2+1/3}$ & $\Delta^{meas}_{2+2/3}$ \\
\hline
A &  88 &  10 &  81 &  27 \\
B & 446 & 120 & 497 & 240 \\
%Ref.\cite{dean08}& 262  & 25 &  & x \\
%Ref.\cite{csa10} & 558  & x & 675 & 384 \\
\end{tabular}
\end{ruledtabular}
\end{table}

The influence of the disorder on the gaps can be understood within the framework of
a widely used phenomenological model \cite{chang83} according to which
the quantized energy levels of the 2DEG are broadened by the disorder 
into bands of localized states of width $\Gamma$. In this model the disorder broadening parameter $\Gamma$
relates the measured and the intrinsic gaps 
\begin{equation}
\Delta^{int}=\Delta^{meas}+\Gamma.
\end{equation}
%  To obtain $\Delta^{int}$ from $\Delta^{meas}$ the
%  disorder broadening parameter $\Gamma$ must be therefore quantitatively known.
This model was instrumental in the analysis of the gaps of the FQHS in the lowest LL
\cite{willett88,du93,mano94,du94} in terms of Laughlin's wavefunction \cite{laughlin} and Jain's CF theory \cite{jainCF}
and we will use it for the FQHSs of the second LL. 

We turn our attention to an independent extraction of $\Gamma$ from the measured data.
As mentioned in the introduction, $\Gamma$ has been estimated from the quantum lifetime $\tau_q$.
The $B$-field dependence of the envelope of the Shubnikov-de Haas oscillations at a fixed temperature contains the
$ \exp \left( -\pi/\omega_C \tau_q \right) $ multiplicative factor from which $\tau_q$ 
and $\Gamma_{SdH}=\hbar/\tau_q$
is extracted \cite{ando74}. Here $\omega_C$ is the cyclotron frequency. 
The values found are summarized in Table II.

%As a further complication, any density inhomogeneities are expected to strongly modify the SdHo and, therefore
%the quantum lifetime and $\Gamma_1$. 

$\Gamma$ can also be found
from a scaling of $\Delta^{meas}$ of the even denominator FQHS at $\nu=5/2$ and $7/2$
with the Coulomb energy $E_C=e^2/\epsilon l_B$ \cite{morf03}. Here $l_B=\sqrt{\hbar/eB}$ is the magnetic length.
Particularly, by assuming that the intrinsic gap of the
$5/2$ and $7/2$ is affected by LLM the same way, $\Delta^{int}=\delta^{int} E_C$ was found 
with the same adimensional intrinsic gap $\delta^{int}$. 
% Strictly speaking the effects of LLM on 5/2 and 7/2 are slightly different as they are observed at different magnetic fields, 
% but the authors argue that the assumption should still be approximately valid \cite{morf03}. 
Therefore $\Gamma_{even}$ is extracted from $\Delta^{meas}=\delta^{int} E_C - \Gamma_{even}$ equation as the
negative intercept of the measured gaps of $5/2$ and $7/2$ FQHS versus $E_C$.
Such an analysis is shown on Fig. 3. As seen in Table II and discussed in Ref.\cite{dean08},
$\Gamma_{even}$ obtained this way may differ significantly from $\Gamma_{SdH}$, by as much as one order of magnitude.

\begin{figure}[t]
% \begin{center}
 \includegraphics[width=1\columnwidth]{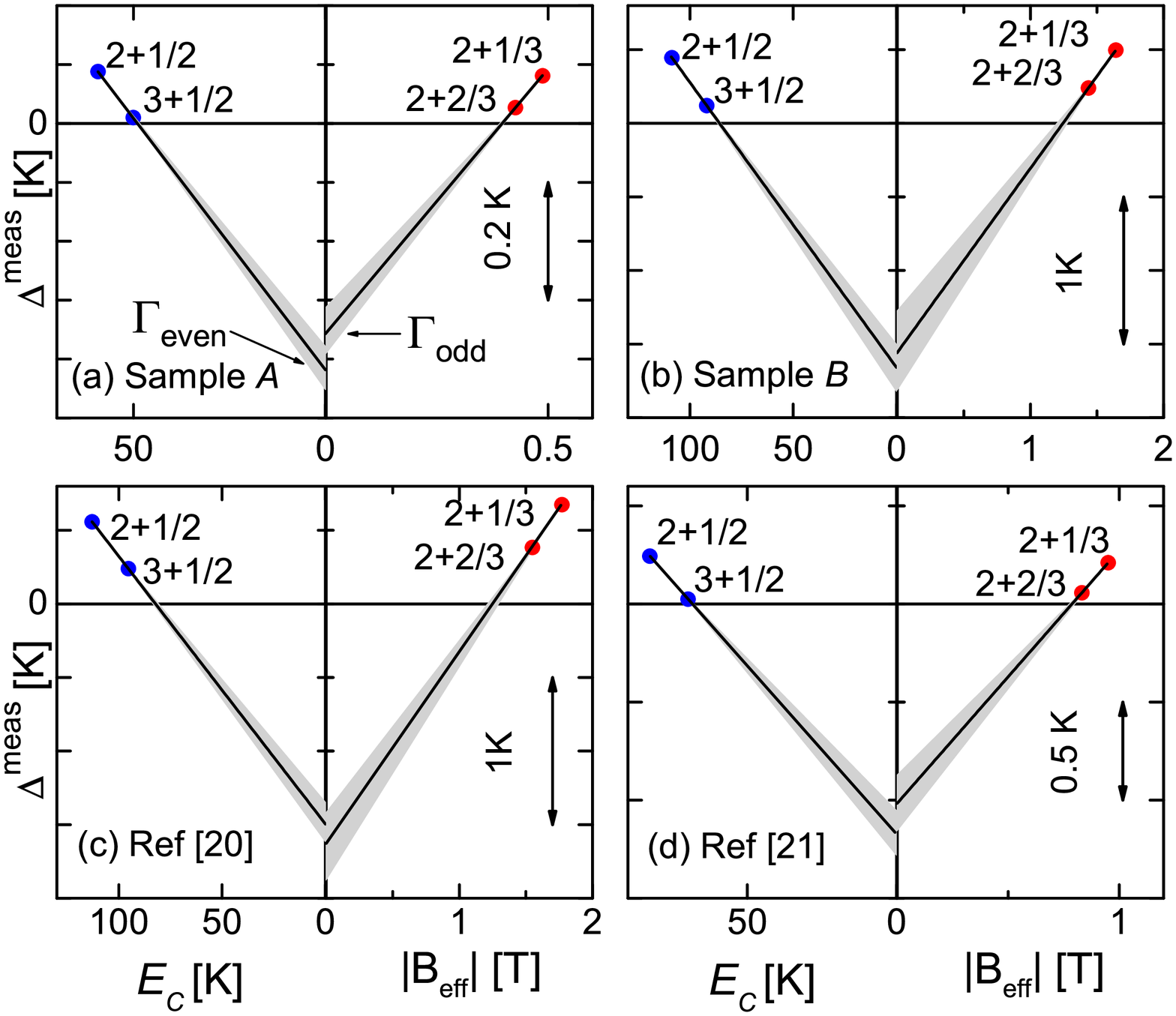}
% \end{center}
 \caption{\label{f3}
Comparison of the two methods of estimating the disorder paramater $\Gamma$ for the four analyzed samples.
The measured gaps of the even denominator FQHS extrapolate to $\Gamma_{even}$ at $E_C=0$,
while those of the $\nu=2+1/3$ and $2+2/3$ FQHS to $\Gamma_{odd}$ at the vanishing absolute value of $B_{eff}$.
The grey shadow is the estimated error for $\Gamma$s.
}
\end{figure}

\begin{table}[b]
\caption{Parameters of samples considered. $n$ is in units of $10^{10}$/cm$^2$, $\Gamma$  
         and $\Delta^{int}_{5/2}$ are in Kelvin.}
\begin{ruledtabular}
\begin{tabular}{l c c c c c c c}
sample & $n$ & $w/l_B$ & $\Gamma_{SdH}$ & $\Gamma_{even}$ & $\Gamma_{odd}$ & $\Delta^{int}_{5/2}$ & $\delta^{int}_{5/2}$\\
\hline
A                & 8.3 & 2.56 & 0.24 & 0.42 & 0.35 & 0.47 & 0.0080 \\
B                & 27.8& 2.52 & 2.04 & 1.65 & 1.55 & 2.04 & 0.019 \\
%% error3
Ref.\cite{csa10} & 30  & 2.61 & 1.55 & 1.50 & 1.62 & 2.12 & 0.019 \\
Ref.\cite{dean08}& 16  & 2.55 & 0.23 & 1.16 & 1.01 & 1.33 & 0.016 \\
\end{tabular}
\end{ruledtabular}
\end{table}

In order to resolve this discrepancy we introduce a third method of extracting $\Gamma$ from the
gaps of the odd denominator states $\nu=2+1/3$ and $2+2/3$. 
%In a recent publication we reported that the  equation $\Delta^{meas}=\hbar e B_{eff}/m_{eff} - \Gamma_{odd}$ 
%which describes the gaps of FQHS in the lowest LL, including those at $\nu=1/3$ and $2/3$, 
%also describes gaps of $\nu=2+1/3$ and $2+2/3$ FQHS with an effective mass $m_{eff}$ 
%independent of the LL index.
Recently we reported that the  equation $\Delta^{meas}=\hbar e |B_{eff}|/m_{eff} - \Gamma_{odd}$ 
describes 
%the gaps of the $\nu=1/3$ and $2/3$ FQHS in the lowest LL as well as those of 
the gaps of the $\nu=2+1/3$ and $2+2/3$ FQHS in the second LL \cite{csa10}.
%with the same effective mass $m_{eff}$ \cite{csa10}.
Here $B_{eff}=5(B-B_{\nu=5/2})$ is the effective magnetic field after 
flux attachment from the CF theory \cite{jainCF,halp93}.
This result was interpreted as being suggestive of Laughlin-correlated $\nu=2+1/3$ and $2+2/3$ FQHS \cite{csa10}.
We use the equation above to extract $\Gamma_{odd}$ for the four analyzed samples.
Fits to the data are shown in Fig. 3. $\Gamma_{odd}$ is the intercept of the fits and the vertical scale 
and its values are listed in Table II.
%
%disorder broadening term obtained appears to be independent of the filling factor.

We found that the disorder broadening terms $\Gamma_{odd}$ and $\Gamma_{even}$ have similar values in each sample.
% This is surprizing given the long range of the extrapolations in Fig. 3.
Typical errors in $\Delta^{meas}$ of $\pm 5$\% for gaps above 100mK and of $\pm 10$\% below 100mK
result in measurement errors in $\Gamma$, shown as a shadow in Fig. 3, of $\pm 12$\%.
We conclude therefore that, within the errors, the even denominator FQHS at $\nu=5/2$ and $7/2$ 
and the two strongest odd denominator FQHS at $2+1/3$ and $2+2/3$ yield {\it the same disorder broadening}
in samples grown in different chambers and covering a wide range of densities and mobilities.
We note that the same disorder broadening for the above FQHSs described by different theories
is possible as they all originate from the same type of CFs. Indeed, the $2+1/3$ and $2+2/3$
FQHS can be understood from motion of flux-two CFs at a finite $B_{eff}$, while
the $5/2$ and $7/2$ are due to paired flux-two CFs at $B_{eff}=0$.
 
$\Gamma_{SdH}$ and $\Gamma_{odd}$ determined from odd denominator FQHS in the second LL are not equal. 
This shows that level broadening is governed by different mechanisms for the low field Shubnikov-de Haas
oscillations and for the high field second LL physics. $\Gamma_{SdH}$ is therefore not expected to be relevant
in determining the intrinsic gaps of FQHS in the second LL, including the $\nu=5/2$ FQHS.
A similar conclusion has also been reached for the FQHS of the lowest LL centered around $\nu=1/2$ \cite{du93,mano94,du94}. 

The experimentally derived $\Delta^{int}_{5/2}$ estimated from Eq.1, together with the corresponding adimensional 
$\delta^{int}_{5/2}=\Delta^{int}_{5/2}/E_C$, 
are found in Table II. For $\Gamma$ we used the average of $\Gamma_{even}$ and $\Gamma_{odd}$. 
%While the intrinsic gap does not contain sample specific disorder effects, it still incorporates
%gap reducing effects of LLM \cite{nuebler10,morf03,wojs06,wojs10} and of finite sample width 
%\cite{nuebler10,morf02,park98,peters08,papic09}. 
The comparison of the experimental and numerically estimated intrinsic gaps
must be performed at the same extent of the LLM \cite{morf03,wojs06,wojs10} and of
finite sample width \cite{park98,morf02,peters08,papic09} as quantified
by the LLM parameter $\kappa=E_C/\hbar \omega_C$ \cite{yoshi84} and adimensional width of the
quantum well $w/l_B$, respectively.
We find that $\delta^{int}_{5/2}=0.019$ listed in Table II. for the sample B and that from Ref.\cite{csa10} 
is only 19\% larger than 0.016, the value calculated from exact diagonalization for similar sample parameters \cite{morf03}. 
Also, $\delta^{int}_{5/2}=$0.0080, 0.019, 0.019, and 0.016 we find in 
samples $A$, $B$, Ref.\cite{csa10}, and Ref.\cite{dean08} compare well with the values
0.014, 0.018, 0.018, and 0.016 we extract from a recent exact diagonalization study \cite{nuebler10}. 
% for a gated 30nm wide quantum well sample at the corresponding densities
We note that, while sample $A$ and that from Ref.\cite{dean08} do not have the same width as 
that in Ref.\cite{nuebler10}, the previous comparison is meaningful
because of the relatively small contributions of finite width effects \cite{nuebler10}.
%the well width of this latter sample is different from those of samples $A$ and of Ref.\cite{dean08}
We conclude that the intrinsic gaps we find are in excellent agreement
with the numerically obtained gaps for 4 samples of very different densities, mobilities, and
which were grown in different MBE chambers.  
These experimental results, when combined with numerical results 
\cite{feinguin08,morf98,morf02,morf03,wojs06,wojs10,park98,peters08,papic09},
strongly support the Pfaffian description of $\nu=5/2$ FQHS.

The data shown in Table II allows us, for the first time, to study the dependence of the intrinsic gap
%% error4
obtained from measurements on LLM. For a meaningful comparison of gaps in
Fig. 4 we plot $\delta^{int}_{5/2}$ as function of the LLM parameter $\kappa$.
The four samples listed in Table II have different widths $w$, but have very similar
%% error5
adimensional widths $w/l_B$ at $\nu=5/2$, and therefore the gap suppression seen in Fig. 4 is solely due to LLM.
We find a decreasing $\delta^{int}_{5/2}$ with an increasing $\kappa$ which is consistent
with expectations \cite{nuebler10,wojs06}. By assuming a linear dependence
for the limited range of $\kappa$ accessed we find $\delta^{int}_{5/2}(\kappa=0)=0.032$ at no LLM.
%%% which is similar to that found for the $\nu=1/3$ FQHS \cite{yoshi84,melik97}
This value compares well with $\approx \! 0.030$, the numerically obtained gap in the ideal 2D limit
\cite{nuebler10,feinguin08,morf98,morf02,morf03}.
%       but it is larger by $\approx \!\! 20$\% than the gap calculated for 8 electrons
%       at the finite width $w/l_B \simeq 2.55$ of our samples \cite{peters08}. 

From our data we also see that $\delta^{int}_{5/2}$ extrapolates to zero at $\kappa_{th}=2.9$ threshold. 
We conclude that the $\nu=5/2$ FQHS should not develop for $\kappa>\kappa_{th}$ or, equivalently, for electron densities 
lower than $n_{th} = 4.4 \times 10^{10} $cm$^{-2}$ even in the limit of no disorder. 
This result could explain the absence of the $\nu=5/2$ FQHS in 2D hole samples 
\cite{manohar94,manfra07,kumar11} in which, due to the enhanced 
effective mass of the holes, values of $\kappa$ lower than 3 have not been achieved.
% is about 5 times larger than that for electron samples of the same density.

Finally we note that the dependence of $\Delta^{meas}_{5/2}$ on the density
in an undoped HIGFET sample has recently been fitted to $\Delta^{meas}_{5/2}=\alpha E_C-\tilde{\Gamma}$,
where $\alpha$ and $\tilde{\Gamma}$  are variables \cite{pan11}. The equation is very similar to the one we used
and one could mistakenly think that $\alpha$ is the intrinsic gap. However, in Ref.\cite{pan11}
$\alpha$ is forced to be a constant of the fit. As discussed earlier and also shown in Fig.4, 
$\delta^{int}_{5/2}$ is a strong function of LLM and, therefore, $E_C$ \cite{nuebler10,wojs06,wojs10,yoshi84}.
The intrinsic gap is therefore expected to change with the density. We conclude that based on the theory
the density-independent constant $\alpha=0.00426$ is {\it not} expected to be the intrinsic gap of the $\nu=5/2$ FQHS 
and that $\tilde{\Gamma}$ is {\it not} the same disorder broadening as $\Gamma$ we found in this present work. 

%An analysis similar to ours of the $\Delta^{meas}_{5/2}$ measured in an undoped HIGFET sample yielded an unusually reduced 
%$\delta^{int}_{5/2}=0.00426$ \cite{pan11}. As we discussed earlier, $\delta^{int}_{5/2}$
%strongly depends on the adimensional sample width $w/l_B$. We point out that while $w/l_B$ is
%roughly constant for the samples we consider, it changes significantly in the work
%in Ref.\cite{pan11} and, therefore, it is not clear whether the fit they used can be applied in their
%situation. The reduced $\delta^{int}_{5/2}$ can also be due to the different
%scattering mechanisms in the doped and undoped samples which is not yet understood \cite{umansky09,gamez11,pan11}.

%For a comparison of experiment and theory it is customary to
%to relate $\delta^{int}_{5/2}$ with the numerically calculated gaps for the Pfaffian
%at specific values of $\kappa$ \cite{nuebler10,wojs06,wojs10}.
%These calculations are often done pertubatively and at the large values of $\kappa$ in sample $A$ 
%it is unclear if such methods apply. We will extrapolate $\delta^{int}_{5/2}$ to $\kappa=0$ instead
%and compare the result of the extrapolation to calculated value in the absence of LLM.

\begin{figure}[t]
% \begin{center}
 \includegraphics[width=0.9\columnwidth]{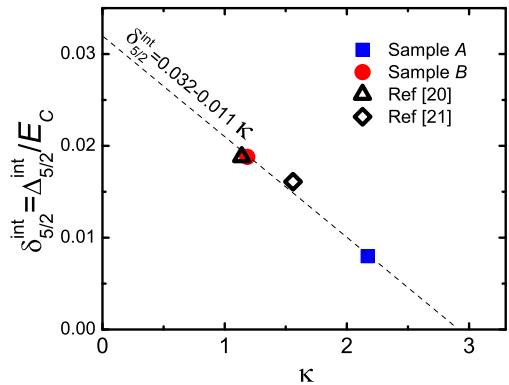}
% \end{center}
 \caption{\label{f4}
 Intrinsic gaps at $\nu=5/2$ as function of the LLM parameter $\kappa$. The dotted line is a linear fit through the data.
}
\end{figure}

In summary, 
%in order to extract the intrinsic gap for the $\nu=5/2$ FQHS 
we have demonstrated that the disorder broadening can be reliably extracted from
$\Delta^{meas}$ of the four major FQHS at $\nu=5/2$, $7/2$, $2+1/3$ and $2+2/3$
for samples over a wide range of densities and grown in different MBE chambers.
The obtained intrinsic gap of the $\nu=5/2$ FQHS
was found to be in an excellent agreement with numerical results lending therefore a
strong support to the Pfaffian description of the $\nu=5/2$ FQHS.
% We have shown that this parameter cannot reliably be extracted from the quantum lifetime and we used two
% different techniques of estimating it based on scaling of the major FQHS of the second LL. 

%\begin{acknowledgments}
N.S. and G.A.C. were supported on 
NSF grant DMR-0907172, M.J.M. acknowledges the Miller Family Foundation, and L.N.P. and K.W.W. 
the Princeton NSF-MRSEC and the Moore Foundation. J.D.W. is supported by a Sandia Laboratories/Purdue University 
Excellence in Science and Engineering Fellowship.

%\end{acknowledgments}

%\bibliography{FQHE}

\end{document}